\documentclass[reprint,twocolumn,secnumarabic,amssymb, nobibnotes, aps, prd, showpacs]{revtex4-1}

\usepackage{graphicx}
\usepackage{dcolumn}
\usepackage{bm}
\usepackage{slashed}
\usepackage{amsfonts, amsmath, amsthm, amssymb} 

\begin{document}

\preprint{AIP/123-QED}

\title{Low energy pion-$\Lambda_b$ interaction}

\author{M. G. L. Nogueira-Santos}
 \email{magwwo@gmail.com}
\author{C. C. Barros, Jr}%
 \email{barros.celso@ufsc.br}
\affiliation{ 
Departamento de F{\'{i}}sica, CFM, Universidade Federal de Santa Catarina\\ Florian{\'{o}}polis SC, CEP 88010-900, Brazil
}%

\date{\today}

\begin{abstract}
In this work we
study the low energy pion-$\Lambda_b$ interaction considering effective chiral Lagrangians 
that include pions, baryons and the corresponding resonances. Interactions
mediated by a $\sigma$ meson exchange are also considered.
The scattering amplitudes are calculated and then we determine the angular 
distributions and polarizations. 
\end{abstract}

\maketitle

 \section{Introduction}

The $b$ baryons physics is a subject that recently have received special attention, as a large
number of this kind of particle is being produced in many experiments, as for example, in the
ATLAS \cite{atl}, LHCb \cite{lhcb} and CMS \cite{cms}. These data in addition with
previous ones \cite{bt1}-\cite{bt4} makes possible a better understanding  of the $b$ baryons
and of their interactions experimentally and at the theoretical level \cite{bth1}-\cite{bth3}.
Quark decays of the type $b\rightarrow s\gamma$ allow the search of physics beyond the standard
model \cite{sm1}-\cite{sm3} and the experimental data impose constraints when new physics is being 
considered.

The polarization and the asymmetry parameter $\alpha$ are fundamental observables that may be
used with this purpose. The basic idea to be considered is the same one that has been used in the study of
hyperons in the HyperCP experiment \cite{hyper1}, \cite{hyper2}, where the $\Lambda$ hyperon
was produced in high energy reactions and its polarization
and the asymmetry parameter $\alpha_\Lambda$ could be determined by the $\Lambda\rightarrow p\pi^-$ 
decay. 
A similar method has been used in the study of $\Lambda_b$ considering the decay
$\Lambda_b\rightarrow J/\psi   \ \Lambda$, produced in high energy collisions, where
$\alpha_b$ and the polarization have been measured \cite{atl}-\cite{cms}. 
If a $b$ or a $s$ quark are produced polarized, it is expected that a large part of this polarization should
remain in the produced baryon, and in general 
a straightforward way to understand these results is to consider that the observed polarization is 
the polarization of the produced hyperon or of the $b$ baryon.
 But in fact, another hypothesis may be considered, that is that the hyperons and the bottom
baryons may be polarized after their production or to have their polarizations altered by final-state interactions.

In \cite{cy}-\cite{ccb2} the polarization of hyperons and antihyperons produced in high energy collisions
have been calculated taking into account the final-state interactions and was successful in explaining the polarization of antihyperons that should be produced unpolarized if this effect was not considered. In the proposed model, based on relativistic hydrodynamics, during the collision a hot expanding medium is formed, and inside this
medium the hyperons are produced. These hyperons interacts with the surrounding particles and then emerges, and may be polarized by these interactions. As far as the most probable particles are pions, 
the dominant interaction is the pion-hyperon interaction at low energies. Despite the fact that
these particles are observed with high energies at the laboratory, the relative energies of the particles inside a
 fluid element is small.

So, in order to understand these interactions a careful study of the low energy pion-hyperon interactions has been done \cite{BH}-\cite{cm}, and recently, in order to improve the model,
the kaon-hyperon interactions has been also studied \cite{sant} and \cite{sant2}. In these models, 
effective nonlinear chiral Lagrangians, including hyperons, resonances and mesons have
been taken into account. Then, the coupling constants, cross-sections and polarizations
have been determined.

Thinking in general terms,
the $\Lambda_b$ polarization may be understood exactly in the same way. In processes where the 
$\Lambda_b$ are considered to be produced unpolarized, they may become polarized if the final-state
interactions are considered, or if they are produced polarized, the final polarization may be altered by these interactions. So, a first step to understand this effect is to study the low energy
pion-$\Lambda_b$ interactions. This study will be conducted in a way analogous to the one that it has been done in 
\cite{BH}-\cite{sant2}, by considering nonlinear effective Lagrangians, and then the $\Lambda_b$ polarization may be studied.

This work will have the following content: in Sec. II the basic formalism will be reviewed, in Sec. III
the $\pi-\Lambda_b$ interaction will be studied, in Sec. IV the results will be presented
and in Sec. V, the conclusions. Some expressions and integrals will be shown in the Appendix.

\section{Basic formalism}

In order to study the low energy $\pi\Lambda_b$ interaction we will
consider a formalism that has been used to describe pion-hyperon 
($\pi Y$) \cite{BH}-\cite{cm} 
and recently the kaon-hyperon ($K Y$)  \cite{sant}, \cite{sant2} interactions, which is based in 
nonlinear
effective chiral Lagrangians that takes into account baryons, resonances and
mesons as degrees of freedom. 
Observing the fact that the low energies pion-nucleon interactions are very well known,
experimentally and theoretically, in \cite{Coelho1983}, \cite{Olsson1975} a model that was successful in 
explaining these interactions has been adapted in 
order to study the $\pi Y$ interactions \cite{BH} (and recently the $KY$ interactions).

In \cite{Coelho1983}, \cite{Olsson1975} the description of the  
$\pi N\rightarrow \pi N$ scattering
have been made considering a chiral model that takes into account vertices with
spin-1/2 and spin-3/2 baryons
that are represented by the effective Lagrangians
\begin{eqnarray}
\label{eq1}
\mathcal{L}_{\pi NN}&=& \frac{g}{2m}\big(\overline{N}\gamma_\mu\gamma_5\vec{\tau}N\big)\cdot\partial^\mu\vec{\phi} \ , \\
\label{eq2}
\mathcal{L}_{\pi N\Delta}&=&   g_\Delta\bigg\{\overline{\Delta}^\mu\Big[g_{\mu\nu}-(Z+1/2)\gamma_\mu\gamma_\nu\Big]\vec{M}N\bigg\}.\partial^\nu\vec{\phi} \ ,  \nonumber
\\
\end{eqnarray}
where $N$, $\Delta$, $\vec{\phi}$ are the nucleon, delta, and pion fields with masses $m$, $m_\Delta$ and $m_\pi$, respectively, $\vec{M}$ and $\vec{\tau}$ are isospin matrices
that combine a nucleon and a delta in a nucleon and  a pion  isospin 1 state, 
and $Z$ is a parameter representing the possibility of the off-shell-$\Delta$ having spin 1/2. The parameters $g$ and $g_\Delta$  are coupling constants. 
So, a reasonable procedure to be followed is to
 adapt these Lagrangians in order to study the $\pi \Lambda_b$ interactions.

The scattering amplitude to the process $\pi \Lambda_b\rightarrow \pi \Lambda_b$ may be 
parametrized as
\begin{equation}
\label{eq3}
 T_{\pi\Lambda_b}=\overline{u}(\vec{p'})\Big[A+\frac{1}{2}(\slashed{k}+\slashed{k}')B\Big]u(\vec{p}),
\end{equation}
where  $u(\vec{p})$ and $\overline{u}(\vec{p'})$ are the spinors that represents
 initial and final $\Lambda_b$, $p_\mu$ and $p_\mu'$ are the initial and final $\Lambda_b$ 
4-momenta and $k_\mu$ $k_\mu'$ and  are the pion initial and final 4-momenta.
$A$ and $B$ are amplitudes that may be calculated.

Thus, in this work
our task will be to calculate the $A$ and $B$ amplitudes for each 
considered diagram.
The scattering matrix is given by
\begin{equation}
M=\frac{T_{\pi\Lambda_b}}{8\pi \sqrt{s}}=f(k,\theta)+ g(k,\theta) \vec{\sigma}.\hat{n},
\end{equation}
which may be decomposed into the spin-non-flip and spin-flip amplitudes $f(k,\theta)$ and $g(k,\theta)$, respectively, written in terms of the variables $k$ and $\theta$,
determined in the center-of-mass frame, where $\theta$ is the scattering angle.
 These spin amplitudes
may be expanded in terms of  partial-waves amplitudes $a_{l\pm}$
by the expressions 
\begin{eqnarray}
 f(k,\theta)&=& \sum_{l=0}^\infty{\Big[(l+1)a_{l+}(k)+la_{l-}(k)\Big] P_l (\theta)}, \\
 g(k,\theta)&=& i\sum_{l=1}^\infty{\Big[a_{l-}(k)-a_{l+}(k)\Big] P_l^{(1)} (\theta)}.
\end{eqnarray}
Using the orthogonality relations of the
 Legendre polynomials, the partial-wave amplitudes may be determined 
\begin{equation}
\label{eq7}
 a_{l\pm}=\frac{1}{2}\int_{-1}^1\Big[P_l(\theta)f_1(k, \theta) +P_{l\pm 1}(\theta)f_2(k,\theta)\Big] d\theta,
\end{equation}
with
\begin{eqnarray}
\label{eq8}
 f_1(k, \theta)&=& \frac{(E+m)}{8\pi \sqrt{s}}[A+(\sqrt{s}-m)B], \\
\label{eq9}
 f_2(k, \theta)&=& \frac{(E-m)}{8\pi \sqrt{s}}[-A+(\sqrt{s}+m)B],
\end{eqnarray}
where $E$ is the $\Lambda_b$ energy and $s$ is a Mandelstam variable (see the Appendix). As
we are interested in studying the low energy interactions ($k<$ 0.5 GeV), a good approximation is to 
consider only the $S$ ($l=0$) and $P$  ($l=1$) waves, as the waves with $l\geq$2 have amplitudes
much smaller than the $S$ and $P$ ones at these energies
 and may be considered just as small corrections. In the following sections these amplitudes will be calculated.

\section{$\pi\Lambda_b$ interaction}
\begin{figure}[!htb]
   \centering
   \includegraphics[width=0.4\textwidth]{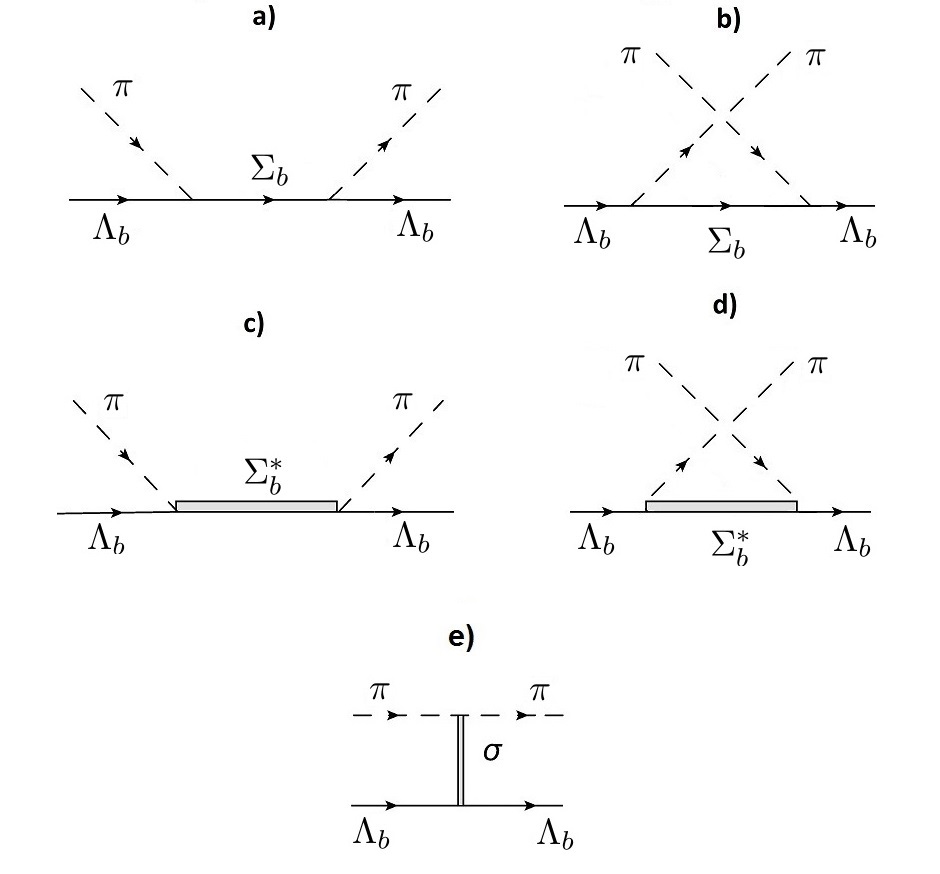}
   \caption{Diagrams to be considered in the $\pi\Lambda_b$ interaction }\label{fg1}
\end{figure}

In this section we will show how to use the formalism presented in the preceding section
in order to compute the amplitudes $A$ and $B$ of eq. (\ref{eq3}) 
considering the Feynman diagrams displayed in Fig. \ref{fg1}.
 The Tab. \ref{tb1} shows the proprieties of the particles that will be considered in the calculations
where $J$ is the spin with parity $\pi$, and $I$, the isospin. 
\begin{table}
\begin{ruledtabular}
\caption{Resonances considered in $\pi\Lambda_b$ interaction}\label{tb1}
\begin{tabular}{cccc}
& $J^\pi$ & $I$ &$Mass$ $(MeV)$ \\
\noalign{\smallskip}\hline\noalign{\smallskip}
$\Sigma_b$  & $1/2^+$&1&5812\\ 
$\Sigma_b^*$  &$3/2^+$&1 &5832\\ 
\end{tabular}
\end{ruledtabular}
\end{table}

The  interaction with a $\Sigma_b$ spin-1/2 baryon in the intermediate state
will be studied
in Sec. \ref{sec31} and the $\Sigma_b^*$ spin-3/2 pole in Sec. \ref{sec32}.
The amplitude for the 
the scalar $\sigma$ meson exchange shown in  Fig. \ref{fg1}$e$, will be
included as a parametrization as it has been made in \cite{BH}-\cite{ccb}. So 
the amplitudes to be considered are  
\begin{eqnarray}
&&A_\sigma=a+bt  \  , \\
\label{eq32}
&&B_\sigma=0   \  ,
\label{eq33}
\end{eqnarray}
with $a=1.05 m_\pi^{-1}$, $b=-0.8m_\pi^{-3}$ where $m_\pi$ is the pion mass \cite{pdg} and $t$ a is the Mandelstam variable defined in the Appendix. More discussions
about these amplitudes may be found in \cite{cm}, \cite{leut1}-\cite{r1}.
From these results the partial-wave amplitudes $f_{l\pm}$ for the diagrams
of Fig. 1 may be calculated.

 
\subsection{$\Sigma_b$ baryon}
\label{sec31}

In order to study the $\pi\Lambda_b$ interactions, the Lagrangians (\ref{eq1}) and (\ref{eq2})
will be adapted considering the pion and the $\Lambda_b$ as initial particles with 
$\Sigma_b$ (spin-1/2) and  $\Sigma_b^*$ (spin-3/2) in the intermediate states. Thus the Lagrangian
(\ref{eq1}) will be written as
\begin{equation}
\label{eq}
\mathcal{L}_{\pi \Lambda_b\Sigma_b}= \frac{g_{\pi \Lambda_b\Sigma_b}}{2m_{\Lambda_b}}\big(\overline{\vec{\Sigma}_b}\gamma_\mu\gamma_5\Lambda_b\big)\cdot\partial^\mu\vec{\phi} \ ,
\end{equation}
which determines the vertex 
\begin{equation}
\label{eq}
V_{\pi \Lambda_b\Sigma_b}= \frac{g_{\pi \Lambda_b\Sigma_b}}{2m_{\Lambda_b}}\slashed{k}\gamma_5\ ,
\end{equation}
where $g_{\pi \Lambda_b\Sigma_b}$ is the coupling constant for the $\pi \Lambda_b\Sigma_b$ interaction and $m_{\Lambda_b}$ is the $\Lambda_b$ mass.

Calculating the contribution of the diagrams $a$ and $b$ of Fig. \ref{fg1} to the
 $T_{\pi\Lambda_b}$ amplitude, relative to the process $\pi\Lambda_b\rightarrow \pi\Lambda_b$
 and writing them in the form of eq.(\ref{eq3}) we have
\begin{eqnarray}
\label{eq14}
A_{\Sigma_b}&=&\frac{g_{\pi \Lambda_b\Sigma_b}^2}{4m_{\Lambda_b}^2}(m_{\Sigma_b}+m_{\Lambda_b})\bigg(\frac{s-m_{\Lambda_b}^2}{s-m_{\Sigma_b}^2}+\frac{u-m_{\Lambda_b}^2}{u-m_{\Sigma_b}^2}\bigg)\ , \nonumber \\
\\
\label{eq15}
B_{\Sigma_b}&=&\frac{g_{\pi \Lambda_b\Sigma_b}^2}{4m_{\Lambda_b}^2}\bigg[\frac{2m_{\Lambda_b}(m_{\Lambda_b}+m_{\Sigma_b})+u-m_{\Lambda_b}^2}{u-m_{\Sigma_b}^2}\nonumber\\
&&-\frac{2m_{\Lambda_b}(m_{\Lambda_b}+m_{\Sigma_b})+s-m_{\Lambda_b}^2}{s-m_{\Sigma_b}^2}\bigg]\ ,
\end{eqnarray}
where
$u$ is a Mandelstam variable and $m_{\Sigma_b}$ is the $\Sigma_b$ mass.
 
Using eq. (\ref{eq14}) and (\ref{eq15}) in eq. (\ref{eq8}) we have
\begin{eqnarray}
\label{eq:}
f_1&=&\frac{g^2_{\pi \Lambda_b\Sigma_b}(E+m_{\Lambda_b})}{32\pi m_{\Lambda_b}^2\sqrt{s}}\bigg\{\frac{1}{s-m_{\Sigma_b}^2}\bigg[(m_{\Sigma_b}+2m_{\Lambda_b}\nonumber\\
&&-\sqrt{s})(s-m_{\Lambda_b}^2)+2m_{\Lambda_b}(m_{\Lambda_b}+m_{\Sigma_b})(m_{\Lambda_b}-\sqrt{s})\bigg]\nonumber\\
&&+\frac{1}{u-m_{\Sigma_b}^2}\bigg[2m_{\Lambda_b}(m_{\Lambda_b}+m_{\Sigma_b})(\sqrt{s}-m_{\Lambda_b})\nonumber\\
&&+(m_{\Sigma_b}+\sqrt{s})(u-m_{\Lambda_b}^2)\bigg]\bigg\}\ .
\end{eqnarray}

Considering eq. (\ref{eqb41}) and (\ref{eqb42}) defined in Appendix, is convenient
write the $f_1$ amplitude as function of $x=\cos \theta$ 
\begin{equation}
\label{eq17}
f_1=\frac{g^2_{\pi \Lambda_b\Sigma_b}(E+m_{\Lambda_b})}{32\pi m_{\Lambda_b}^2\sqrt{s}}\bigg\{a_1+\frac{b_1+c_1 x}{\gamma_{\Sigma_b}+2|\vec{k}|^2x}\bigg\} \ ,
\end{equation}
where
\begin{eqnarray}
a_1&=&\tfrac{(m_{\Sigma_b}+2m_{\Lambda_b}-\sqrt{s})(s-m_{\Lambda_b}^2)+2m_{\Lambda_b}(m_{\Lambda_b}+m_{\Sigma_b})(m_{\Lambda_b}-\sqrt{s})}{s-m_{\Sigma_b}^2};\nonumber\\
b_1&=&2m_{\Lambda_b}(m_{\Lambda_b}+m_{\Sigma_b})(m_{\Lambda_b}-\sqrt{s})\nonumber\\
&&+(m_{\Sigma_b}+\sqrt{s})(2Ek_0-m_\pi^2);\nonumber\\
c_1&=&2|\vec{k}|^2(\sqrt{s}+m_{\Sigma_b});\nonumber\\
\gamma_{\Sigma_b}&=&m_{\Sigma_b}^2-m_{\Lambda_b}^2-m_\pi^2+2Ek_0.\nonumber
\end{eqnarray}
and $k_0$ is the pion energy.
 
From (\ref{eq14}), (\ref{eq15}) and (\ref{eq9}), we have
\begin{eqnarray}
\label{eq18}
f_2&=&\frac{g^2_{\pi \Lambda_b\Sigma_b}(E-m_{\Lambda_b})}{32\pi m_{\Lambda_b}^2\sqrt{s}}\bigg\{\frac{1}{s-m_{\Sigma_b}^2}\bigg[(m_{\Sigma_b}+2m_{\Lambda_b}+\sqrt{s})(m_{\Lambda_b}^2-s)\nonumber\\
&&-2m_{\Lambda_b}(m_{\Lambda_b}+m_{\Sigma_b})(m_{\Lambda_b}+\sqrt{s})\bigg]\nonumber\\
&&+\frac{1}{u-m_{\Sigma_b}^2}\bigg[2m_{\Lambda_b}(m_{\Lambda_b}+m_{\Sigma_b})(m_{\Lambda_b}+\sqrt{s})\nonumber\\
&&+(\sqrt{s}-m_{\Sigma_b})(u-m_{\Lambda_b}^2)\bigg]\bigg\}\ ,
\end{eqnarray}
and following the same procedure that has been followed in the calculation of $f_1$ we obtain
\begin{equation}
\label{eq19}
f_2=\frac{g^2_{\pi \Lambda_b\Sigma_b}(E-m_{\Lambda_b})}{32\pi m_{\Lambda_b}^2\sqrt{s}}\bigg\{a_2+\frac{b_2+c_2 x}{\gamma_{\Sigma_b}+2|\vec{k}|^2x}\bigg\}\ ,
\end{equation}
where
\begin{eqnarray}
a_2&=&\tfrac{(m_{\Sigma_b}+2m_{\Lambda_b}+\sqrt{s})(m_{\Lambda_b}^2-s)-2m_{\Lambda_b}(m_{\Lambda_b}+m_{\Sigma_b})(m_{\Lambda_b}+\sqrt{s})}{s-m_{\Sigma_b}^2};\nonumber\\
b_2&=&-2m_{\Lambda_b}(m_{\Lambda_b}+m_{\Sigma_b})(\sqrt{s}+m_{\Lambda_b})+(\sqrt{s}-m_{\Sigma_b})(2Ek_0-m_\pi^2);\nonumber\\
c_2&=&2|\vec{k}|^2(\sqrt{s}-m_{\Sigma_b}).\nonumber
\end{eqnarray}

Thus, using eqs. (\ref{eq17}) and (\ref{eq19}) in eq. (\ref{eq7})
 the $S$, $P_1$ and $P_3$ partial-waves amplitudes may be calculated 
\begin{eqnarray}
\label{eq:}
a_S^{({\Sigma_b})}&=&\frac{g^2_{\pi \Lambda_b\Sigma_b}}{64\pi m_{\Lambda_b}^2 \sqrt{s}}\bigg\{2(E+m_{\Lambda_b})a_1+(E+m_{\Lambda_b})b_1 I_0\nonumber\\
&&+\Big[(E+m_{\Lambda_b})c_1+(E-m_{\Lambda_b})b_2\Big]I_1\nonumber\\
&&  +(E-m_{\Lambda_b})c_2I_2\bigg\}\ ,\\
\label{eq:}
a_{P_1}^{({\Sigma_b})}&=&\frac{g^2_{\pi \Lambda_b\Sigma_b}}{64\pi m_{\Lambda_b}^2 \sqrt{s}}\bigg\{2(E-m_{\Lambda_b})a_2+(E-m_{\Lambda_b})b_2 I_0\nonumber\\
&&+\Big[(E+m_{\Lambda_b})b_1+(E-m_{\Lambda_b})c_2\Big]I_1\nonumber\\
&&  +(E+m_{\Lambda_b})c_1I_2\bigg\}\ ,\\
\label{eq:}
a_{P_3}^{({\Sigma_b})}&=&\frac{g^2_{\pi \Lambda_b\Sigma_b}}{64\pi m_{\Lambda_b}^2 \sqrt{s}}\bigg\{\Big(\frac{m_{\Lambda_b}-E}{2}\Big)b_2 I_0\nonumber\\
&&+\Big[(E+m_{\Lambda_b})b_1+\Big(\frac{m_{\Lambda_b}-E}{2}\Big)c_2\Big]I_1 \nonumber\\
&&+\Big[\frac{3}{2}(E-m_{\Lambda_b})b_2+(E+m_{\Lambda_b})c_1\Big]I_2\nonumber\\ 
&&+\frac{3}{2}(E-m_{\Lambda_b})c_2I_3  \bigg\}\ ,
\end{eqnarray}
where $I_0$, $I_1$, $I_2$ and $I_3$ are integrals defined in the Appendix.

These amplitudes will be used in Sec. \ref{sec4} 
in order to calculate the observables of interest.

\subsection{$\Sigma_b^*$ baryon}
\label{sec32}

Now,
adapting the Lagrangian (\ref{eq2}) in order to describe a
the spin-3/2 $\Sigma_b^*$ resonance in the intermediate state, 
 represented in the diagrams $c$ and $d$ of  Fig.\ref{fg1}, 
 we have
\begin{equation}
\label{eq}
\mathcal{L}_{\pi\Lambda_b{\Sigma^*_b}}=  g_{\pi\Lambda_b\Sigma^*_b}\Big\{\overline{{\vec{\Sigma}_b}^{*\mu}}\big[g_{\mu\nu}-(Z+1/2)\gamma_\mu\gamma_\nu\big]\Lambda_b\Big\}\cdot\partial^\nu\vec{\phi}\ ,
\end{equation}
as the $\Lambda_b$ hyperon has isospin 0, the isospin matrix $\vec{M}$ in eq. (\ref{eq2})
is not needed. So, the Feynman vertex is 
\begin{equation}
\label{eq}
V_{\pi\Lambda_b{\Sigma^*_b}}=  g_{\pi\Lambda_b{\Sigma^*_b}}\Big[k^\mu-(Z+1/2)\gamma^\mu\slashed{k}\Big]\ ,
\end{equation}
and then the amplitudes are
\begin{eqnarray}
\label{eq25}
A_{\Sigma^*_b}&=&\frac{2g_{\pi\Lambda_b{\Sigma^*_b}}^2}{9m_{\Lambda_b}}\bigg[\frac{\nu_{\Sigma^*_b}}{\nu_{\Sigma^*_b}^2-\nu^2}\hat{A}+m_{\Lambda_b}(-a_0+a_z k.k')\bigg]\ , \nonumber\\
\\
\label{eq26}
B_{\Sigma^*_b}&=&\frac{2g_{\pi\Lambda_b{\Sigma^*_b}}^2}{9m_{\Lambda_b}}\bigg[\frac{\nu}{\nu_{\Sigma^*_b}^2-\nu^2}\hat{B}-(2m_{\Lambda_b}^2b_0)\nu\bigg]\ ,
\end{eqnarray}
where $g_{\pi\Lambda_b\Sigma^*_b}$ is the coupling constant and $\nu$, $\nu_{\Sigma_b^*}$ and $k.k'$ are defined in the Appendix. The other terms that appear in eq. (\ref{eq25}) and (\ref{eq26})
are defined as  
\begin{eqnarray}
\hat{A}&=&\frac{(m_{\Sigma^*_b}+m_{\Lambda_b})^2-m_\pi^2}{2m_{\Sigma^*_b}^2}\bigg[2m_{\Sigma^*_b}^3-2m_{\Lambda_b}^3-2m_{\Lambda_b}m_{\Sigma^*_b}^2\nonumber\\
&&-2m_{\Lambda_b}^2m_{\Sigma^*_b}+m_\pi^2(2m_{\Lambda_b}-m_{\Sigma^*_b})\bigg]\nonumber\\
&&+ \frac{3}{2}(m_{\Lambda_b}+m_{\Sigma^*_b})t\ ,
\label{eq:}
\end{eqnarray}
\begin{eqnarray}
\hat{B}&=&\frac{1}{2m_{\Sigma^*_b}^2}\bigg[(m_{\Sigma^*_b}^2-m_{\Lambda_b}^2)^2\nonumber\\
&&-2m_{\Lambda_b}m_{\Sigma^*_b}(m_{\Lambda_b}+m_{\Sigma^*_b})^2-2m_\pi^2(m_{\Lambda_b}+m_{\Sigma^*_b})^2\nonumber\\
&&+6m_\pi^2m_{\Sigma^*_b}(m_{\Lambda_b}+m_{\Sigma^*_b})+m_\pi^4\bigg]+ \frac{3}{2}t\ ,
\label{eq:}
\end{eqnarray}
\begin{eqnarray}
a_0&=&\frac{(m_{\Lambda_b} +m_{\Sigma^*_b})}{m_{\Sigma^*_b}^2}(2m_{\Sigma^*_b}^2+m_{\Lambda_b} m_{\Sigma^*_b}-m_{\Lambda_b} ^2+2m_\pi^2)\ ,\nonumber\\
\\
\label{eq:}
a_z&=&\frac{4}{m_{\Sigma^*_b}^2}\Big[(m_{\Sigma^*_b}+m_{\Lambda_b})Z+(2m_{\Sigma^*_b}+m_{\Lambda_b})Z^2\Big]\ , \\
\label{eq:}
b_0&=&\frac{4Z^2}{m_{\Sigma^*_b}^2}\ ,
\label{eq:}
\end{eqnarray}
where $m_{\Sigma_b^*}$ is the resonance mass.

Then, substituting (\ref{eq25}) and (\ref{eq26}) in the eq. (\ref{eq8}), we find  
\begin{eqnarray}
f_1&=&\frac{g_{\pi\Lambda_b{\Sigma^*_b}}^2}{9}\frac{(E+m_{\Lambda_b})}{4\pi \sqrt{s}}\bigg\{ \frac{\hat{A}+(\sqrt{s}-m_{\Lambda_b})\hat{B}}{m_{\Sigma^*_b}^2-s}\nonumber\\
&&+ \frac{\hat{A}+(m_{\Lambda_b}-\sqrt{s})\hat{B}}{m_{\Sigma^*_b}^2-u}-a_0+a_zkk'\nonumber\\
&&-2m_{\Lambda_b} (\sqrt{s}-m_{\Lambda_b})b_0\nu \bigg\}\ ,
\label{eq:}
\end{eqnarray}

\noindent
that in terms of $x$ may be written as
\begin{equation}
\label{eq:}
f_1=\frac{g_{\pi\Lambda_b{\Sigma^*_b}}^2(E+m_{\Lambda_b})}{36\pi \sqrt{s}}\bigg\{\alpha_1+\beta_1 x+ \frac{a_1+b_1 x}{\gamma_{\Sigma^*_b}+2|\vec{k}|^2x}\bigg\}\ ,
\end{equation}
where
\begin{eqnarray}
\alpha_1&=&\frac{3(\sqrt{s}+m_{\Sigma^*_b})}{4m_{\Sigma^*_b}^2s}\Big[m_{\Sigma^*_b}^2s-(m_{\Lambda_b}^2-m_\pi^2)^2\Big]+ \frac{(E_{\Sigma^*_b}+m_{\Lambda_b})^2}{\sqrt{s}+m_{\Sigma^*_b}}\nonumber\\
&&-a_0+a_zk_0^2+b_0(m_{\Lambda_b}-\sqrt{s})(2Ek_0+|\vec{k}|^2);\nonumber\\
\beta_1&=&\frac{3|\vec{k}|^2}{m_{\Sigma^*_b}-\sqrt{s}}-\Big[a_z-(m_{\Lambda_b}-\sqrt{s})b_0\Big]|\vec{k}|^2;\nonumber\\
a_1&=&3(2m_{\Lambda_b}+m_{\Sigma^*_b}-\sqrt{s})(|\vec{q}_{\Sigma^*_b}|^2-|\vec{k}|^2)\nonumber\\
&&+(m_{\Sigma^*_b}-2m_{\Lambda_b}+\sqrt{s})(E_{\Sigma^*_b}+m_{\Lambda_b})^2;\nonumber\\
b_1&=&3|\vec{k}|^2(2m_{\Lambda_b}+m_{\Sigma^*_b}-\sqrt{s});\nonumber\\
\gamma_{\Sigma_b^*}&=&m_{\Sigma^*_b}^2-m_{\Lambda_b}^2-m_\pi^2+2Ek_0.\nonumber
\end{eqnarray}

For the $f_2$ amplitude we have
\begin{eqnarray}
f_2&=&\frac{g_{\pi\Lambda_b{\Sigma^*_b}}^2}{9}\frac{(E-m_{\Lambda_b})}{4\pi \sqrt{s}}\bigg\{\frac{-\hat{A}+(\sqrt{s}+m_{\Lambda_b})\hat{B}}{m_{\Sigma^*_b}^2-s}\nonumber\\
&&-\frac{\hat{A}+(\sqrt{s}+m_{\Lambda_b})\hat{B}}{m_{\Sigma^*_b}^2-u}+a_0-a_zkk'\nonumber\\
&&-2m_{\Lambda_b}(\sqrt{s}+m_{\Lambda_b})b_0\nu \bigg\}\ ,
\label{eq:}
\end{eqnarray}
then,
\begin{equation}
\label{eq:}
f_2=\frac{g_{\pi\Lambda_b{\Sigma^*_b}}^2(E-m_{\Lambda_b})}{36\pi \sqrt{s}}\bigg\{\alpha_2+\beta_2 x-\frac{a_2+b_2 x}{\gamma_{\Sigma^*_b}+2|\vec{k}|^2x}\bigg\} \ ,
\end{equation}
where
\begin{eqnarray}
\label{eq:}
\alpha_2&=&-3 \frac{|\vec{q}_{\Sigma^*_b}|^2-|\vec{k}|^2}{\sqrt{s}+m_{\Sigma^*_b}}+\frac{(E_{\Sigma^*_b}+m_{\Lambda_b})^2}{\sqrt{s}-m_{\Sigma^*_b}}\nonumber\\
&&+a_0-a_zk_0^2-b_0(\sqrt{s}+m_{\Lambda_b})(2Ek_0+|\vec{k}|^2);\nonumber\\
\beta_2&=&\frac{-3|\vec{k}|^2}{m_{\Sigma^*_b}+\sqrt{s}}+\Big[a_z-(m_{\Lambda_b}+\sqrt{s})b_0\Big]|\vec{k}|^2;\nonumber\\
a_2&=&3(2m_{\Lambda_b}+m_{\Sigma^*_b}+\sqrt{s})(|\vec{q}_{\Sigma^*_b}|^2-|\vec{k}|^2)\nonumber\\
&&+(m_{\Sigma^*_b}-2m_{\Lambda_b}-\sqrt{s})(E_{\Sigma^*_b}+m_{\Lambda_b})^2;\nonumber\\
b_2&=&3|\vec{k}|^2(2m_{\Lambda_b}+m_{\Sigma^*_b}+\sqrt{s})\nonumber\ .
\end{eqnarray}
Finally, the resulting partial-waves $S$ and $P$ amplitudes are
\begin{eqnarray}
 a_S^{({\Sigma_b^*})}&=&\frac{g^2_{\pi\Lambda_b{\Sigma^*_b}}}{72\pi \sqrt{s}}\bigg\{2(E+m_{\Lambda_b})\alpha_1+\frac{2}{3}(E-m_{\Lambda_b})\beta_2\nonumber\\
&&+(E+m_{\Lambda_b})a_1 I_0+\Big[(E+m_{\Lambda_b})b_1-(E-m_{\Lambda_b})a_2\Big]I_1\nonumber\\
&&  -(E-m_{\Lambda_b})b_2I_2\bigg\}\ ,\\
\label{eq:}
a_{P_1}^{({\Sigma_b^*})}&=&\frac{g^2_{\pi\Lambda_b{\Sigma^*_b}}}{72\pi  \sqrt{s}}\bigg\{2(E-m_{\Lambda_b})\alpha_2+ \frac{2}{3}(E+m_{\Lambda_b})\beta_1\nonumber\\
&&-(E-m_{\Lambda_b})a_2 I_0+\Big[(E+m_{\Lambda_b})a_1-(E-m_{\Lambda_b})b_2\Big]I_1 \nonumber\\
&& +(E+m_{\Lambda_b})b_1I_2\bigg\}\ ,\\
\label{eq:}
a_{P_3}^{({\Sigma_b^*})}&=&\frac{g^2_{\pi\Lambda_b{\Sigma^*_b}}}{72\pi  \sqrt{s}}\bigg\{ \frac{2}{3}(E+m_{\Lambda_b})\beta_1-\Big(\frac{m_{\Lambda_b}-E}{2}\Big)a_2 I_0\nonumber\\
&&+\Big[(E+m_{\Lambda_b})a_1-\Big(\frac{m_{\Lambda_b}-E}{2}\Big)b_2\Big]I_1\nonumber\\ 
  &&+\Big[\frac{3}{2}(m_{\Lambda_b}-E)a_2+(E+m_{\Lambda_b})b_1\Big]I_2\nonumber\\  
	 &&+\frac{3}{2}(m_{\Lambda_b}-E)b_2I_3  \bigg\}\ .
\label{eq:}
\end{eqnarray}

\section{Results}
\label{sec4}

Considering the amplitudes obtained before, now
it is possible to calculate the observables of interest, that is the 
subject of this section. First of all we must observe that the calculated
 partial-wave amplitudes are real, and then violate the unitarity of the $S$ matrix.
 These amplitudes may be unitarized \cite{BH}-\cite{cm} with     
\begin{equation}
a_{l\pm}^U=\frac{a_\pm}{1-i|\vec{k}|a_\pm}\ .
\end{equation}

 In the center-of-mass frame
 the differential cross section is given by
\begin{equation}
\frac{d\sigma}{d\Omega}=|f|^2+|g|^2\ ,
\label{eq:}
\end{equation}
\noindent
and 
the total cross section may be obtained by
integrating this expression over the solid angle that results in
\begin{equation}
\sigma_T=4\pi \sum_l{\Big[(l+1)|a_{l+}^U|^2+l|a_{l-}^U|^2\Big]}\ .
\end{equation}

The polarization is defined by
\begin{equation}
\vec{P}=-2 \frac{{\rm Im}(f^*g)}{|f|^2+|g|^2}\hat{n}\  ,
\label{eq:}
\end{equation}

\noindent
where $\hat n$ is a unitary vector normal to the scattering plane and the phase shifts 
 are given by
\begin{equation}
\delta_{l\pm}=\tan^{-1}(|\vec{k}|a_\pm)\ .
\label{eq:}
\end{equation}

\begin{figure}[!htb]
   \centering
   \includegraphics[width=0.5\textwidth]{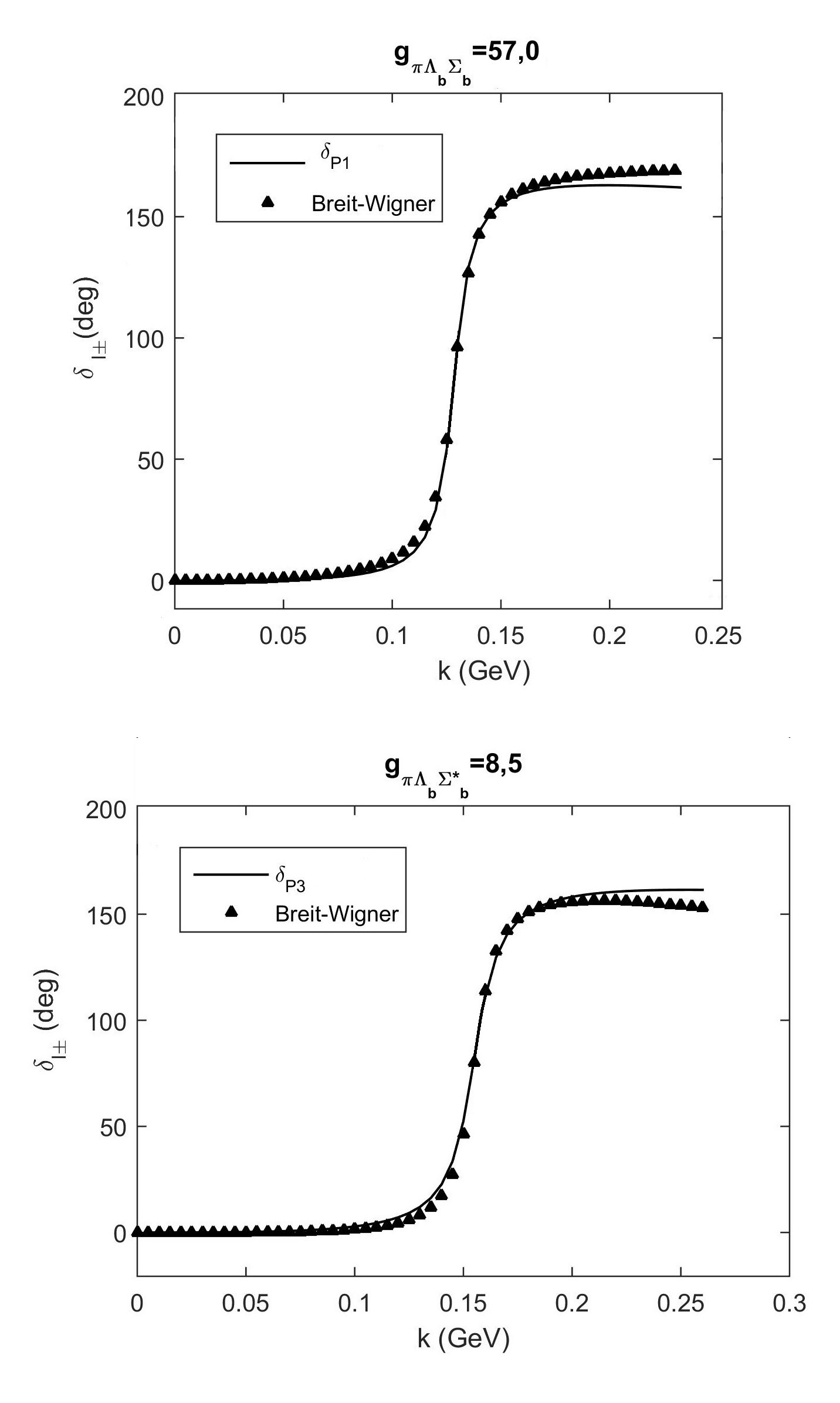}
   \caption{Breit-Wigner fit for the $\Sigma_b$ and $\Sigma_b^*$ resonances }\label{figBW}
\end{figure}


The parameters used in the calculations are shown in Tab. \ref{tb1}  and \ref{tb2}. The coupling
constants have been determined by comparing our results for the phase-shifts with the Breit-Wigner
expression (see the Appendix) \cite{BH}, \cite{sant}, as it is shown in Fig. \ref{figBW}. 
In Fig. 3, 4, 5 and 6 the total cross section, phase-shifts, differential cross section and polarization are
plotted as functions of the pion momentum $k=|\vec{k}|$  and $x$.
The gaps between the lines in Fig. 6, for
the polarization, are 10 $MeV$.

\begin{table}
\caption{Constants of $\pi\Lambda_b$ interaction }\label{tb2}
\begin{ruledtabular}
\begin{tabular}{ll}
$m_{\Lambda_b}$ & $5620$ $MeV$ \\
$g_{\pi\Lambda_b\Sigma_b}$ & $63.0$ $GeV^{-1}$ \\
$g_{\pi\Lambda_b\Sigma_b^*}$ & $8.5$ $GeV^{-1}$ \\
$Z$ & -0.5 \\
\end{tabular}
\end{ruledtabular}
\end{table}
\begin{figure}[!htb]
   \centering
   \includegraphics[width=0.5\textwidth]{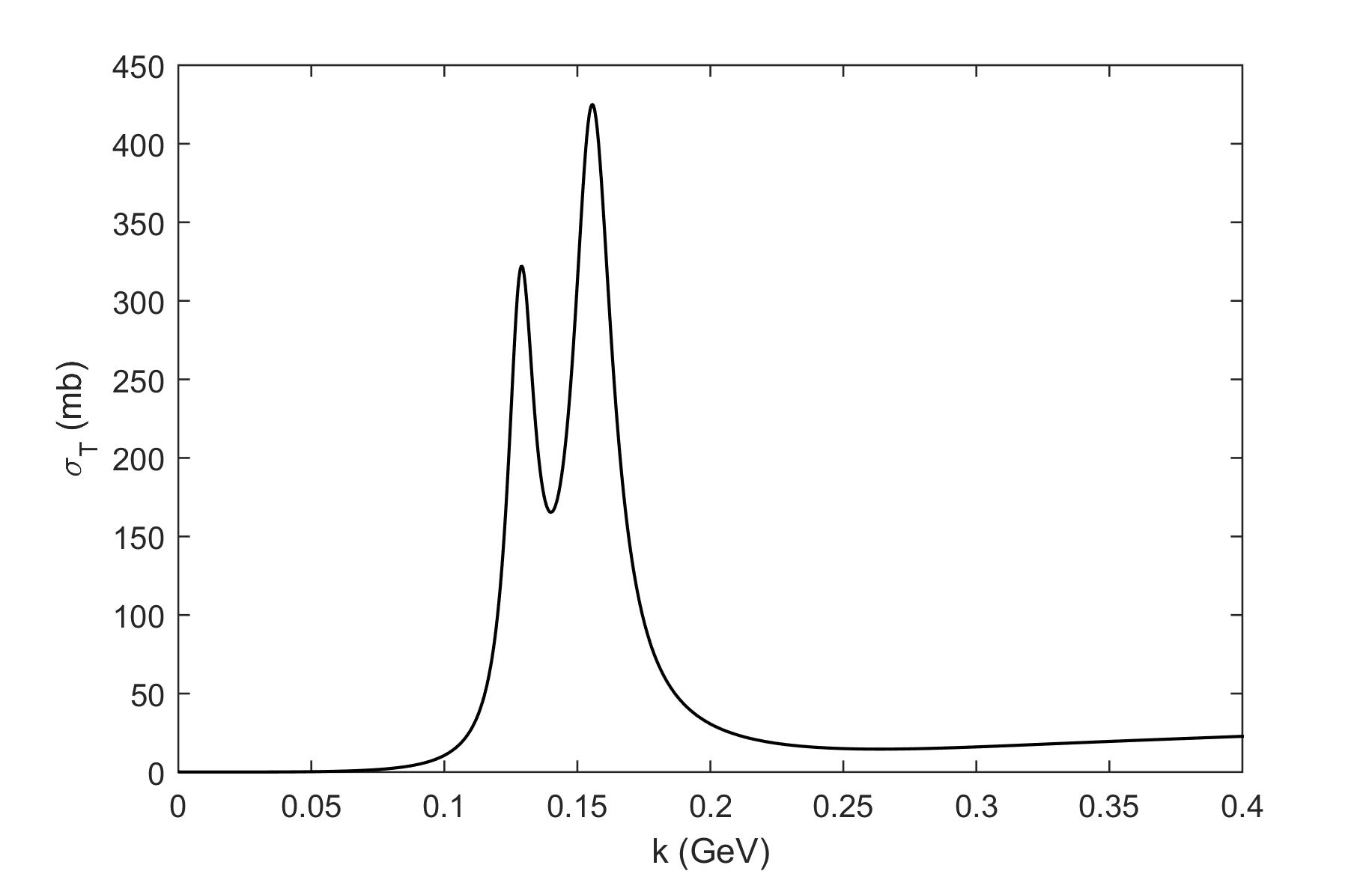}
   \caption{Total Cross Section of the $\pi\Lambda_b$ interaction }\label{fig2}
\end{figure}
\begin{figure}[!htb]
   \centering
   \includegraphics[width=0.5\textwidth]{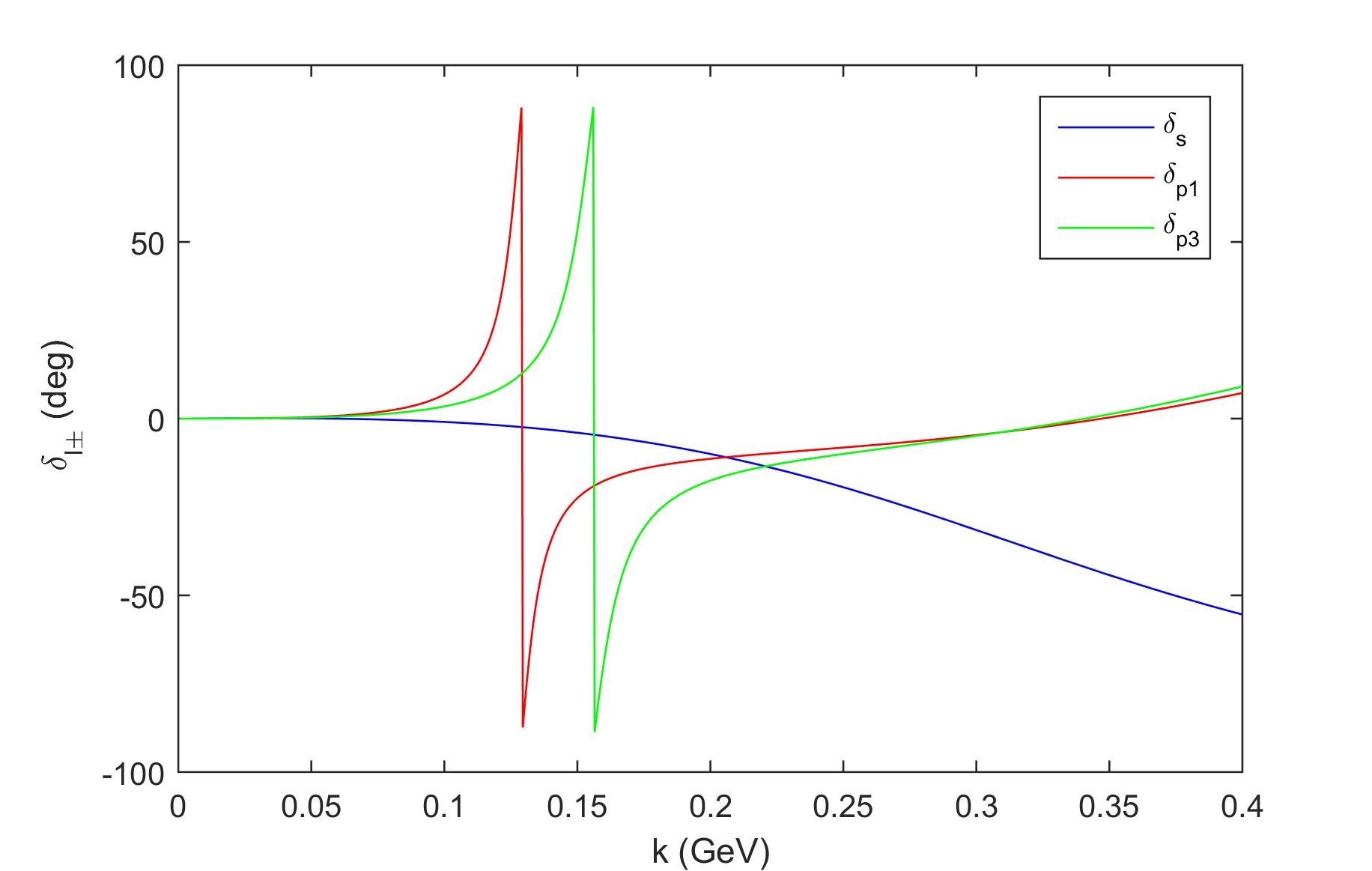}
   \caption{Phase Shifts of the $\pi\Lambda_b$ interaction }\label{fig3}
\end{figure}
\begin{figure}[!htb]
   \centering
   \includegraphics[width=0.5\textwidth]{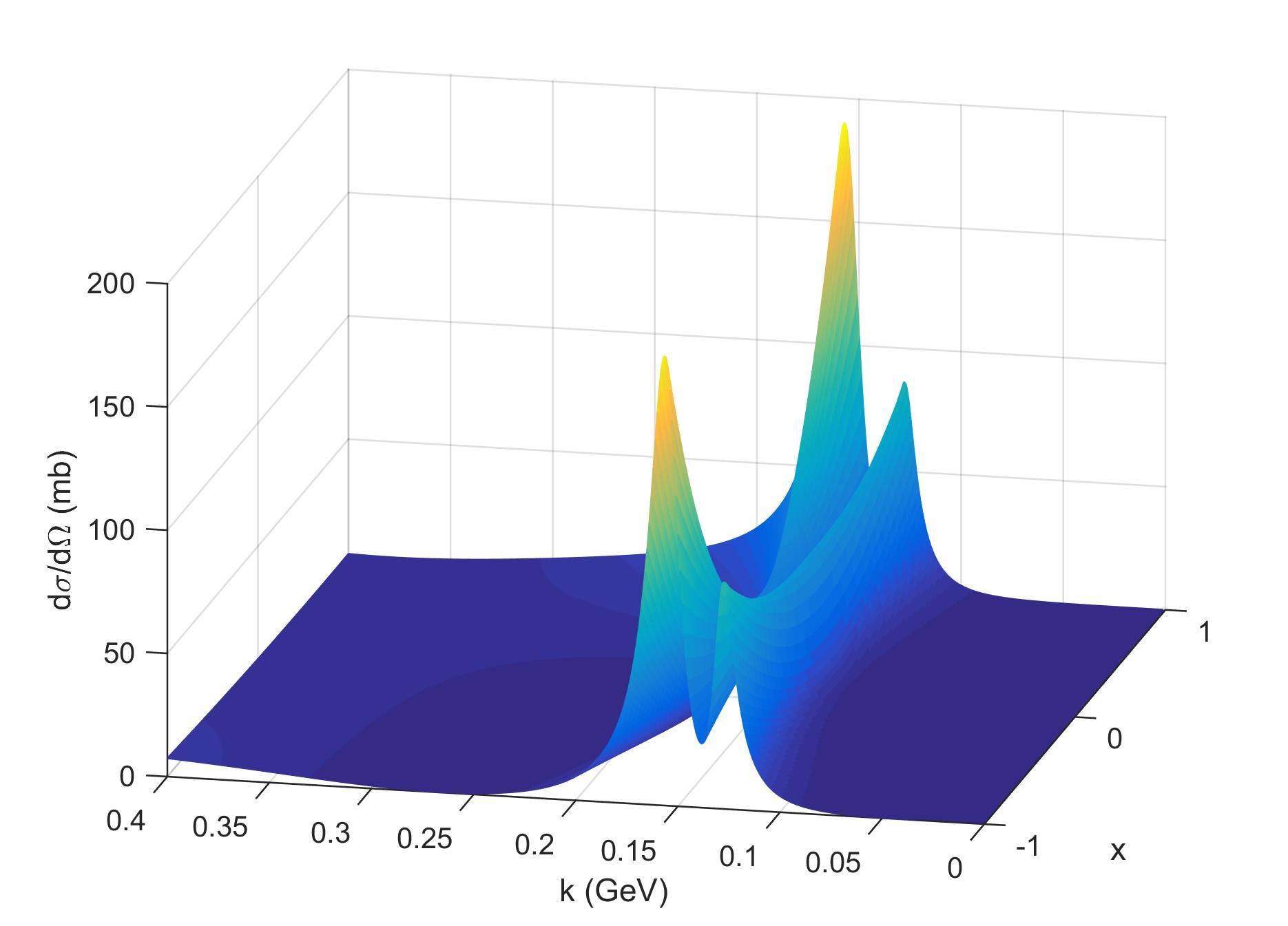}
   \caption{Differential Cross Section  of the $\pi\Lambda_b$ interaction }\label{fig4}
\end{figure}
\begin{figure}[!htb]
   \centering
   \includegraphics[width=0.5\textwidth]{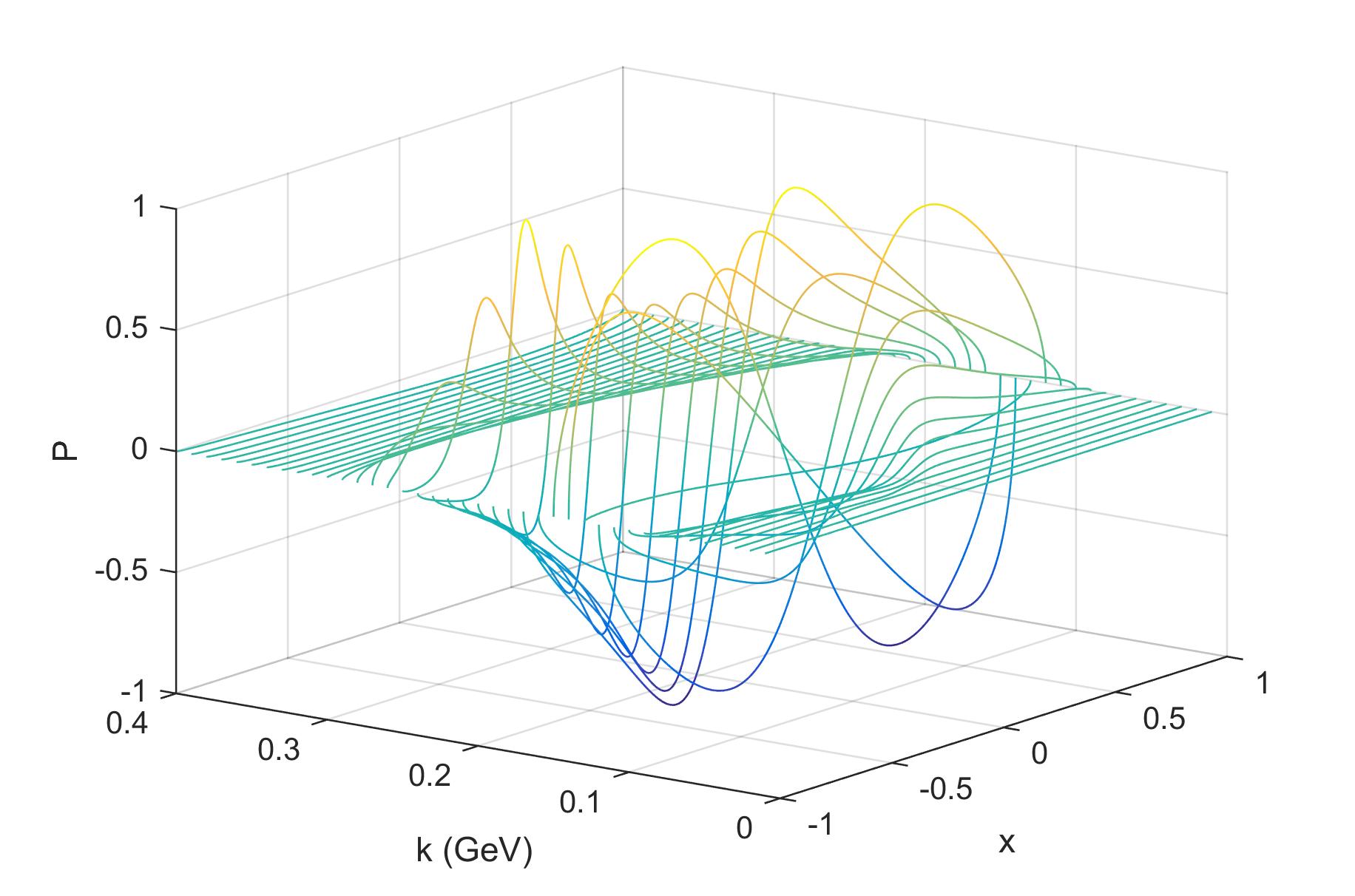}
   \caption{Polarization in the $\pi\Lambda_b$ interaction }\label{fig5}
\end{figure}

\section{Conclusions}

In this work the low energy pion-$\Lambda_b$ has been studied. We considered Lagrangians
that include pions, $\Lambda_b$, $\Sigma_b$ and $\Sigma_b^*$. In the interaction mediated
by $\sigma$, a parametrization has been considered. The coupling constants have been determined
and then the phase-shifts, cross-sections and polarizations have been calculated. As it should be
expected, at low energies, the resonances dominate the cross sections, as it may be seen
in Fig. 3 and 5. This behavior is similar to the one observed in pion-hyperon \cite{BH}-\cite{cm}
and kaon-hyperon \cite{sant}, \cite{sant2} interactions and is very well determined in the pion-nucleon
interactions, where the $\Delta$ particle dominates the cross section in the spin-3/2 and 
isospin-3/2 channel.

At the LHCb experiment \cite{lhcb}, the transverse polarization of $\Lambda_b$ at $\sqrt{s}$= 7 TeV has been measured 0.06$\pm$0.07$\pm$0.02 and in the CMS \cite{cms}, for $\sqrt{s}$ =7 and 8 TeV,
the measured polarization was 0.00$\pm$0.06$\pm$0.06, that are small values. Observing Fig.
6, we may see that in general, the polarization is not so large, except near the $\Sigma_b$
and   $\Sigma_b^*$ masses.
If this polarization is considered in mechanisms such as the ones presented in
\cite{cy}-\cite{ccb2},  that take into account
 the final-state interactions of the $\Lambda_b$, a relatively small
value of the polarization $1\%-2\%$ (or even smaller) should remain when the average is computed.
 Obviously in order to  determine the exact
value, the calculations must be made, that is a task for a future work.
\begin{figure}[!htb]
   \centering
   \includegraphics[width=0.4\textwidth]{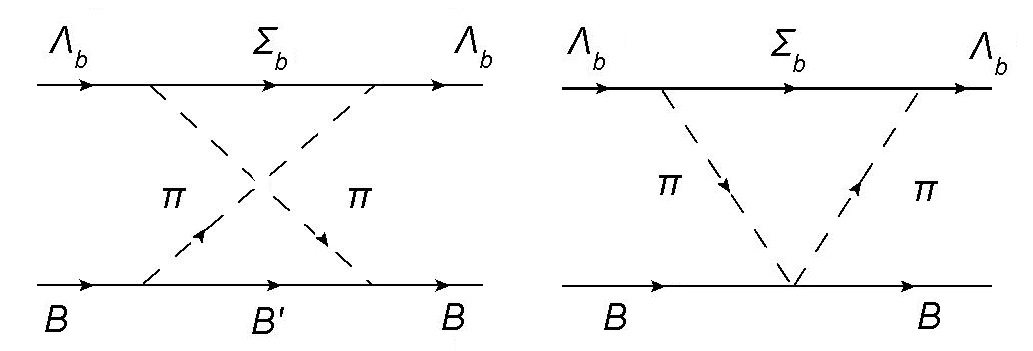}
   \caption{Examples of two pions exchange in the $Y\Lambda_b$ or $N\Lambda_b$ interactions }\label{fig6}
\end{figure}

The results of this work are also important in the determination of nucleon-$\Lambda_b$ and hyperon-$\Lambda_b$ potentials, as the pion-$\Lambda_b$ interaction is needed in order to calculate diagrames of the type of Fig. 
\ref{fig6}. This kind of potential is fundamental in many physical systems, as  for example
in order to describe exotic
nuclei that contain bottom baryons. Then, the model presented in this work deals with
an important aspect of physics, where still there are much to be done, that is
describing and understanding  the interactions and proprieties
of the heavy-quark baryons.


\section{Acknowledgments}

This study has been partially supported by the Coordena\c c\~ao
de Aperfei\c coamento de Pessoal de N\'{\i}vel Superior
(CAPES) – Finance Code 001.

\section{Appendix}
\subsection{Kinematics Relations}
Considering a process where
 $p_\mu$ and $p_\mu'$ are
the initial and final baryon four-momenta, $k_\mu$ and $k_\mu'$ are the initial
 and final meson 
four-momenta, the Mandelstam variables may be written in
terms of variables defined in the center-of-mass frame, 
$|\vec{k}|$ is the three-momentum absolute value
 of the incident particle and  $\theta$, the scattering angle

\[
s=(p+k)^2=(p'+k')^2=m^2_{\Lambda_b}+m_\pi^2+2Ek_0-2\vec{k}.\vec{p} \ ,
\]
\begin{equation}
\label{eqb41}
 s=m_{\Lambda_b}^2+m_\pi^2+2Ek_0+2{|\vec{k}|} ^2\ ,
\end{equation}
\[
u=(p'-k)^2=(p-k')^2=m^2_{\Lambda_b}+m_\pi^2-2Ek_0-2\vec{k}'.\vec{p}\ ,
\]
\begin{equation}
\label{eqb42}
 u=m^2_{\Lambda_b}+m_\pi^2-2{|\vec{k}|}^2x-2Ek_0\ ,
\end{equation}
\[
t=(p-p')^2=(k-k')^2=2|\vec{k}|^2x-2|\vec{k}|^2  \   .
\]
\begin{equation}
\label{eqb43}
 t=2|{\vec{k}}| ^2(x-1)\ ,
\end{equation}
where
\begin{equation}
\label{eq47}
x=\cos\theta\ ,
\end{equation}
and the energies are defined as
\begin{eqnarray}
&&k_0=k'_0=\sqrt{|\vec{k}|^2+m_\pi^2}\ , \\
&&E=E'=\sqrt{|\vec{k}|^2+m^2_{\Lambda_b}}\ ,
\end{eqnarray}
also
\begin{equation}
E+k_0=\sqrt{s} \ ,
\label{eq:}
\end{equation}
is the total energy of the system.

Other variables of interest are
\begin{eqnarray}
&&\nu_{\Sigma_b^*}=\frac{m_{\Sigma_b^*}^2-m^2_{\Lambda_b}-k.k'}{2m_{\Lambda_b}}\ , \\
&&\nu=\frac{s-u}{4m_{\Lambda_b}}=\frac{2Ek_0+|\vec{k}|^2+|\vec{k}|^2x}{2m_{\Lambda_b}}\ , \\
&&k.k'=m_\pi^2+|\vec{k}|^2-|\vec{k}|^2x=k_0^2-|\vec{k}|^2x\ ,
\end{eqnarray}
where $m_{\Lambda_b}$, $m_{\Sigma_b^*}$ and $m_\pi$ are the $\Lambda_b$ mass, the resonance  $\Sigma_b^*$  mass and the pion mass, respectively.
We also define the relations
\begin{eqnarray}
&&(E_{\Sigma_b^*}\pm m_{\Lambda_b})=\frac{(m_{\Sigma_b^*}\pm m_{\Lambda_b})^2-m_\pi^2}{2m_{\Sigma_b^*}}\ , \\
&&(\vec{q}_{\Sigma_b^*})^2=E_{\Sigma_b^*}^2- m_{\Lambda_b}^2=(E_{\Sigma_b^*}+ m_{\Lambda_b})(E_{\Sigma_b^*}- m_{\Lambda_b})\ , \nonumber \\
\end{eqnarray}
where $E_{\Sigma_b^*}$ and $\vec{q}_{\Sigma_b^*}$  are the energy and the and the 3-momentum of 
the $\Sigma_b^*$ baryon that is produced in the
intermediate states. 
\subsection{Integrals}
Some integrals that appear in the calculations has the general form
\begin{equation}
I_n=\int_{-1}^1{\frac{x^ndx}{\gamma+2k^2x}}\ ,
\label{eq:}
\end{equation}
where $\gamma$ is defined in the text, $x$ is given by
eq. (\ref{eq47}), $k=|\vec{k}|$ and $n$ is an integer.  
Thus, we have 
\begin{eqnarray}
I_0&=&\int_{-1}^1{\frac{dx}{\gamma+2k^2x}}=\frac{1}{2k^2}\ln\bigg(\frac{\gamma+2k^2}{\gamma-2k^2}\bigg)\ ,\\
I_1&=&\int_{-1}^1{\frac{x dx}{\gamma+2k^2x}}\\
&=&\frac{2}{2k^2}- \frac{\gamma}{(2k^2)^2}\ln\bigg(\frac{\gamma+2k^2}{\gamma-2k^2}\bigg)\ , \\
I_2&=&\int_{-1}^1{\frac{x^2 dx}{\gamma+2k^2x}}\nonumber\\
&=&-\frac{2\gamma}{(2k^2)^2}+\frac{\gamma^2}{(2k^2)^3}\ln\bigg(\frac{\gamma+2k^2}{\gamma-2k^2}\bigg)\ , \\
I_3&=&\int_{-1}^1{\frac{x^3 dx}{\gamma+2k^2x}}\nonumber\\
&=&\frac{2(2k^2)^2+6\gamma^2}{3(2k^2)^3}
-\frac{\gamma^3}{(2k^2)^4}\ln\bigg(\frac{\gamma+2k^2}{\gamma-2k^2}\bigg)\ .
\end{eqnarray}
\subsection{Breit-Wigner expression}
 The relativistic Breit-Wigner expression is determined in terms of experimental quantities
\begin{equation}
\delta_{l\pm}=\tan^{-1}\Bigg[\frac{\Gamma_0\Big(\frac{|\vec{k}|}{|\vec{k}_0|}\Big)^{2J+1}}{2(m_r-\sqrt{s})}\Bigg]\ ,
\label{eq19}
\end{equation}
where $\Gamma_0$ is the width, $|\vec{k}_0|$ is the  momentum at the peak of the 
resonance in the
center-of-mass system,
$m_r$ is its mass and $J$ the total angular momentum (spin) of the resonance.


\end{document}